\documentstyle[amsmath,epsf,epsfig,preprint,aps,amssymb]{revtex}

\begin{document}
\preprint{SNUTP 04-022}

\title{\Large \bf Gauges in the Bulk II: Models with
Bulk Scalars}
\author{Jihn E.
Kim$^{(a)}$\footnote{jekim@phyp.snu.ac.kr}, Gary B.
Tupper$^{(b)}$, and Raoul D. Viollier
$^{(b)}$\footnote{viollier@science.uct.ac.za} }
\address{
$^{(a)}$School of Physics and Center for Theoretical Physics,
Seoul National University, Seoul 151-747,
Korea\\
$^{(b)}$Institute of Theoretical Physics and Astrophysics,
Department of Physics, University of Cape Town, Private Bag,
Rondebosch 7701, South Africa}

 \maketitle

\begin{abstract}
Extending previous work in Randall-Sundrum type models,
we construct low-energy effective actions for braneworlds with
a bulk scalar field, with special attention to the case of
BPS branes. Holding the branes at fixed coordinate position
with a general ansatz for the bulk metric, and imposing the
Einstein frame as a gauge condition, we obtain a scalar-tensor
theory with only one scalar degree of freedom related to the proper
brane separation. The formalism is applicable even when there is
direct coupling of the bulk scalar and brane matter, as in the
Horava-Witten model. We further show that the usual moduli space
approximation actually describes a non-BPS three-brane system.

\noindent [Key words: Braneworld cosmology, Radion, Horava-Witten
model]
\end{abstract}

%Pacs 11.25.Mj\\
\newpage

\newcommand{\bea}{\begin{eqnarray}}
\newcommand{\eea}{\end{eqnarray}}
\def\beq{\begin{equation}}
\def\eeq{\end{equation}}

\def\one{\bf 1}
\def\two{\bf 2}
\def\five{\bf 5}
\def\ten{\bf 10}
\def\tenb{\overline{\bf 10}}
\def\fiveb{\overline{\bf 5}}
\def\threeb{{\bf\overline{3}}}
\def\three{{\bf 3}}
\def\fb{{\overline{F}\,}}
\def\hb{{\overline{h}}}
\def\Hb{{\overline{H}\,}}

\def\To{${\cal T}_1$\ }
\def\Tt{${\cal T}_2$\ }
\def\Z{${\cal Z}$\ }
\def\ot{\otimes}
\def\tr{{\rm tr}}
\def\sinw{{\sin^2 \theta_W}}

\def\slash#1{#1\!\!\!\!\!\!/}

\newcommand{\dis}[1]{\begin{equation}\begin{split}#1\end{split}\end{equation}}
\newcommand{\beqa}[1]{\begin{eqnarray}#1\end{eqnarray}}

\def\be{\begin{equation}}
\def\ee{\end{equation}}
\def\ben{\begin{enumerate}}
\def\een{\end{enumerate}}
\def\lsl{ l \hspace{-0.45 em}/}
\def\ksl{ k \hspace{-0.45 em}/}
\def\qsl{ q \hspace{-0.45 em}/}
\def\psl{ p \hspace{-0.45 em}/}
\def\ppsl{ p' \hspace{-0.70 em}/}
\def\dsl{ \partial \hspace{-0.45 em}/}
\def\Dsl{ D \hspace{-0.55 em}/}
\def\matrix{ \left(\begin{array} \end{array} \right) }

\def\ma{m_A}
\def\mf{m_f}
\def\mz{m_Z}
\def\mw{m_W}
\def\ml{m_l}
\def\ms{m_S}
\def\dag{\dagger}

 \def\NCA{{\em Nuovo Cimento} }
 \def\NIM{{\em Nucl. Instrum. Methods} }
 \def\NIMA{{\em Nucl. Instrum. Methods} A }
 \def\NP{{\em Nucl. Phys.} }
 \def\NPB{{\em Nucl. Phys.} B }
 \def\PL{{\em Phys. Lett.} }
 \def\PLB{{\em Phys. Lett.} B }
 \def\PRL{{\em Phys. Rev. Lett.} }
 \def\PRD{{\em Phys. Rev.} D }
 \def\PR{{\em Phys. Rev.} }
 \def\RMP{{\em Rev. Mod. Phys.} }
 \def\ZPC{{\em Z. Phys.} C }
 \def\PHYSICA{{\em Physica} D }
 \def\CMP{{\em Commun. Math. Phys.} }
 \def\PREP{{\em Phys. Rep.} }
 \def\JMP{{\em J. Math. Phys.} }
 \def\CQG{{\em Class. Quant. Grav.} }
 \def\ANN{{\em Annals of Physics} }
 \def\APP{{\em Acta Phys. Polon.} }
 \def\RPP{{\em Rep. Prog. Phys.} }

\section{Introduction}\label{sec:Int}

Braneworld models have become the focus of
intense theoretical activity in
the last few years (for recent reviews see
\cite{Maa1}). Much of the
attention has been triggered by the
Randall-Sundrum (RS) models
\cite{RS2}
with a purely anti-de-Sitter bulk. However, to stabilize the brane
separation, and hence the hierarchy solution of the first RS model,
Goldberger and Wise
\cite{Gold3}
introduced a massive bulk scalar together with brane
potentials. Bulk scalar models have also been suggested to
alleviate the cosmological constant problem
\cite{Ark4}
(see, however
\cite{Fors5}),
and for driving inflation
\cite{Koba6}.
Thus, phenomenological considerations lead away from the RS
models and towards something closer to the five-dimensional reduction
\cite{Luka7}
of Horava-Witten $M$-theory
\cite{Hora8} that inspired them.

Among the models involving a bulk scalar $\Phi$ one readily
distinguishes two extreme cases. The first of these occurs when
the bulk (brane) potential $U (\Phi)$ ($V (\Phi)$) is dominated
by the bulk cosmological constant (brane tension)
with small residual potentials $u(\Phi)$ ($v(\Phi)$).
The problem of obtaining a four-dimensional low-energy effective
action for such a situation has been addressed by Kanno and Soda
\cite{Kann9}. Indeed, up to Kaluza-Klein corrections, their two-brane
effective action readily obtains by replacing $\Phi (x,y)$ with its
zero mode $\eta (x)$ and integrating over the fifth coordinate
$x^{5} = y$ using the first metric ansatz of Chiba
\cite{Chib10}. This follows because at leading order in the
low-energy expansion the bulk scalar plays no role in determining the
bulk geometry which is identical to that of the RS model.

The opposite extreme occurs when the bulk scalar controls the bulk
geometry. In the case that the potentials $U (\Phi)$ and $V (\Phi)$
derive from a superpotential $W (\Phi)$ the solution of the static
vacuum geometry reduces to a set of first-order BPS-like equations
\cite{Dewo11}.
The Horava-Witten model exemplifies this category.
An additional feature of the Horava-Witten model is that $V (\Phi)$ is
the volume modulus of the Calabi-Yau space, hence the scalar directly
couples to matter, and in particular with the inclusion of
nonrelativistic matter static bulk solutions do not exist
\cite{Arno12}.
%\\[.2cm]
In the restricted case of no $\Phi$-matter coupling, a low-energy
effective action for BPS braneworlds has been given in
\cite{Brax13,Palm14} using the moduli space approximation. The
moduli space approximation proceeds from the static vacuum
solution by replacing the Minkowski metric $\eta_{\mu \nu}$ on the
brane with $g_{\mu \nu}\;(x)$, where $x^{\mu}$ are coordinates
tangential to the brane, and promoting the coordinate orthogonal
to the positive/negative tension brane to a field $X^{(\pm)5}
(x)$; the result is a bi-scalar-tensor theory. Clearly one scalar
corresponds to a relative displacement between the branes, however
the second scalar represents a centre-of-mass displacement that is
spurious on a two-brane orbifold. Indeed, perturbation theory
evidences a single scalar mode for BPS branes \cite{Brax15}. We
will say more on this point anon. While the original moduli space
approximation based on moving branes in a fixed background cannot
be used in the interesting case of the Horava-Witten model due to
the direct $\Phi$-matter coupling, the alternate formulation of
Palma and Davis \cite{Palm14} can. Then one is led to a remarkable
conclusion: the Horava-Witten model is cosmologically excluded due
to the centre-of-mass mode \cite{Brax13}.

In this paper we pursue the low-energy effective action for
BPS braneworlds from a different approach which extends our
previous treatment
\cite{Kim16} of RS type models. Specifically,
we maintain the branes at fixed coordinate $x^{5} = y$, while taking a
rather general ansatz for the five-dimensional metric that includes
the graviton zero mode $g_{\mu \nu} (x)$. The other metric functions,
and now the scalar $\Phi$, are restricted by imposing the $\mu$-5 bulk
Einstein equation. A residual freedom is fixed by requiring that the
resulting effective action be in the Einstein frame.
There is no centre-of-mass mode in this two-brane system.
For an exponential superpotential the effective action can be given
in closed form, and the Horava-Witten model appears as a particular case.

The remainder of this paper is organized as follows: in Section
\ref{sec:BPS}, we briefly review the construction of the static
vacuum solution following \cite{Dewo11}. Then, in Section
\ref{sec:Gau}, we present our metric ansatz and analyze the
constraints on the metric functions. Section \ref{sec:Ein} gives
our construction of the effective action in the Einstein gauge
with the exponential superpotential as an example. In Section
\ref{sec:Jor} we discuss the Jordan gauge analogous to
\cite{Chib10} as well as the moduli space gauge and show that the
latter actually describes a different non-BPS three-brane system.
Conclusions are presented in Section \ref{sec:Con}. An appendix
gives the effective action for RS-type models.
%%%%%%%%%%%%%%%%%%%%%%
%%%%%%%%%%%%%%%%%%%%%%
%\newpage
\section{Vacuum BPS Branes}\label{sec:BPS}
We begin with the action $S = S_{\rm bulk} + S_{\rm brane}$, with
\begin{equation}
\displaystyle{
S_{\rm bulk} = \frac{1}{K_{(5)}} \; \int \; d^{5} x \sqrt{g_{(5)}} \;
\left[ - \frac{1}{2} \; R_{(5)} + \frac{1}{2} \; g_{(5)}^{MN}
 \; \Phi_{, M} \; \Phi_{, N} - U (\Phi) \right] \; \; ,
}
\end{equation}
\begin{equation}
\displaystyle{
S_{\rm brane} = - \frac{1}{K_{(5)}} \; \int \; d^{4} x
 \sqrt{- g_{(4)}} \; V (\Phi) \; \; . }
\end{equation}
Only one brane is shown, the
second being introduced below; throughout the Gibbons-Hawking
surface terms are left implicit. In vacuum, the metric and bulk
scalar are chosen as the $x^{\mu}$ independent forms
\vspace{-0.25cm}
\begin{equation}
\displaystyle{
d S_{(5)}^{2} = {\rm e}^{- 2 A (y)} \; \eta_{\mu \nu} \;
d x^{\mu} \; d x^{\nu} - d y^{2} \; , \; \Phi = \Phi (y) \; \; .
}
\end{equation}
The nonvanishing components of the Einstein tensor are
\vspace{-0.25cm}
\begin{equation}
\displaystyle{
G_{(5) \mu \nu} = \eta_{\mu \nu} \; {\rm e}^{- 2 A}
 \; \left[ 3 A'' - 6 (A')^{2} \right] }
\end{equation}
\vspace{-0.75cm}
\begin{equation}
\displaystyle{
G_{(5) 55} = 6 (A')^{2} }
\end{equation}
and those of the bulk energy-momentum tensor are
\begin{equation}
\displaystyle{
K_{(5)} \; T_{(5) \mu \nu}^{(\Phi)} = \eta_{\mu \nu} \; {\rm e}^{- 2 A} \;
\left[ \frac{1}{2} \; \left( \Phi' \right)^{2} + U (\Phi) \right]  }
\end{equation}
%%%EQ 7
\begin{equation}
\displaystyle{
K_{(5)} \; T_{(5) \mu \nu}^{(\Phi)} = \frac{1}{2}
\; \left( \Phi' \right)^{2} - U (\Phi) }
\end{equation}
where prime denotes derivative with respect to $y$.
The positive tension brane will be placed at $y = 0$ so that
the Einstein equations are
%%%EQ 8
\begin{equation}
\displaystyle{
{\rm e}^{- 2A} \; \left[ 3 A'' - 6 \left( A' \right)^{2} \right]
 = {\rm e}^{- 2 A} \;
\left[ \frac{1}{2} \; \left( \Phi' \right)^{2} + U (\Phi )
 \right] + {\rm e}^{- 2 A_{0}} \;
V \left( \Phi_{0} \right) \; \delta (y)  }
\end{equation}
%%%EQUATION 9
\begin{equation}
\displaystyle{
6 \; \left( A' \right)^{2} = \frac{1}{2} \; \left( \Phi '
\right)^{2} - U ( \Phi ) }
\end{equation}
with $A_{0} = A (0), \Phi_{0} = \Phi (0)$. The $\Phi$ field equation is
%%%EQUATION 10
\begin{equation}
\displaystyle{
\Phi '' - 4 \; A' \; \Phi ' =
\frac{\partial U ( \Phi)}{\partial \Phi} +
\frac{\partial V ( \Phi)}{\partial \Phi} \; \delta \; (y) \; \; . }
\end{equation}
Note it is more convenient to recast Eq. (8), by multiplying with
${\rm e}^{2 A}$, as
%%%EQUATION 11
\begin{equation}
\displaystyle{
3 \; A'' - 6 \left( A' \right)^{2} = \frac{1}{2} \;
\left( \Phi ' \right)^{2} +
U ( \phi ) + V \left( \Phi_{0} \right) \; \delta  (y)  }
\end{equation}
which combines with Eq. (9) to yield
%%%EQUATION 12
\vspace{-0.25cm}
\begin{equation}
\displaystyle{
3 \; A'' = \left( \Phi ' \right)^{2} + V \left( \Phi_{0} \right)
 \; \delta (y) \; \; . }
\end{equation}
Integrating Eqs. (12) and (10) around $y = 0$ gives
%%EQUATION 13
\vspace{-0.25cm}
\begin{equation}
\displaystyle{6 \; A' (0) = V \left( \Phi_{0} \right) }
\end{equation}
%%EQUATION 14
\begin{equation}
\displaystyle{
2 \; \Phi ' (0) = \frac{\partial V
\left( \Phi_{0} \right) }{ \partial \Phi_{0} } }
\end{equation}
where orbifold symmetry has been assumed.
%\\[.2cm]
Given $U (\Phi)$ in terms of a superpotential $W (\Phi)$,
%%%EQUATION 15
\vspace{0.5cm}
\begin{equation}
\displaystyle{
U ( \Phi) = \frac{1}{8} \; \left[ \frac{ \partial W
(\Phi)} {\partial \Phi} \right]^{2} -
\frac{1}{6} \; \left[ W (\Phi ) \right]^{2} \; \; , }
\end{equation}
one readily verifies that the Einstein equations (11, 12)
and scalar field equation (10) are satisfied away from the
brane if
%%%EQUATION 16
\vspace{-0.5cm}
\begin{equation}
\displaystyle{
A' = \frac{1}{6} \; W ( \Phi ) }
\end{equation}
%%%EQUATION 17
\begin{equation}
\displaystyle{
\Phi ' = \frac{1}{2} \; \frac{\partial W (\Phi) }{
\partial \Phi } \; \; .}
\end{equation}
In addition, the boundary conditions of Eqs. (13, 14) are
satisfied if
%%%EQUATION 18
\begin{equation}
\displaystyle{
W \left( \Phi_{0} \right) = V \left( \Phi_{0} \right)
\; \; \; , \; \; \;
\frac{ \partial W \left( \Phi_{0} \right) }{ \partial \Phi_{0} } =
\frac{ \partial V \left( \Phi_{0} \right) }{ \partial \Phi_{0} }
\; \; , }
\end{equation}
i.e. $W \left( \Phi_{0} \right)$ is tangent to $V \left( \Phi_{0}
\right)$. When $W ( \Phi) = V ( \Phi)$ the boundary conditions are
automatically fulfilled by the solution of Eqs. (16,17), and the
brane is BPS. As an example, for
%\vspace{-0.5cm}
%%%EQUATION 19
\vspace{-0.5cm}
\begin{equation}
\displaystyle{
W ( \Phi ) = V ( \Phi ) = W_{0} \; {\rm e}^{\alpha
( \Phi_{0} - \Phi ) } \; \; ,
}
\end{equation}
%%%EQUATION 20
\vspace{-1cm}
\begin{equation}
\displaystyle{
\frac{d \Phi}{d y} = - \frac{\alpha}{2} \; W_{0} \;
{\rm e}^{\alpha ( \Phi_{0} - \Phi )}
}
\end{equation}
integrating to
\vspace{-0.5cm}
%%%EQUATION 21
\vspace{0.5cm}
\begin{equation}
\displaystyle{
\Phi (y) = \Phi_{0} + \frac{1}{\alpha} \; \mbox{ln} \;
\left( 1 - \frac{\alpha^{2}}{2} \; W_{0} \; y \right) \; \; .
}
\end{equation}
Then
%%%EQUATION 22
\begin{equation}
\displaystyle{
\frac{d A}{d y} = \frac{1}{6} \; \frac{W_{0}}{1 -
\frac{\alpha^{2}}{2} \; W_{0} \; y}
}
\end{equation}
so that
%%%EQUATION 23
\vspace{-0.5cm}
\begin{equation}
\displaystyle{
A (y) = A_{0} - \frac{1}{3 \alpha^{2}} \; \mbox{ln} \; \left( 1 -
\frac{\alpha^{2}}{2} \; W_{0} y \right) \; \; .
}
\end{equation}
The integration constant will be chosen as $A_{0} = 0$, hence
%%%EQUATION 24
\begin{equation}
\displaystyle{
{\rm e}^{- A(y)} = \left( 1 - \frac{\alpha^{2}}{2} \; W_{0} \; y
 \right)^{\frac{1}{3\alpha^2}} \; \; .
}
\end{equation}
Clearly the limit $\alpha \rightarrow 0$ reproduces the RS model
\cite{RS2}, including the fine-tuning of the bulk cosmological
constant and brane tension, if we set $W_{0} = 6 k$ where $k$ is
the AdS$_{5}$ curvature. The Horava-Witten model corresponds to
$\alpha = \sqrt{2}$, and the self-tuning model \cite{Ark4} to
$\alpha = 2/\sqrt{3}$. Note, however, using Eqs. (5, 22), $G_{(5)
55}$ possesses a bulk singularity. To avoid the naked singularity
it is necessary to add a second brane at the other orbifold fixed
point $y = \ell < 2/ \alpha^{2} \; W_{0}$ with
%%%EQUATION 25
\begin{equation}
\displaystyle{
V \left( \Phi_{\ell} \right) = - W \left( \Phi_{\ell} \right), \;
\frac{   \partial V \left( \Phi_{\ell} \right) }
     {   \partial \Phi_{\ell}                  } =
\frac{ - \partial W \left( \Phi_{\ell} \right) }
     {   \partial \Phi_{\ell}                  } \; \; .
}
\end{equation}
It is this additional fine-tuning that undermines the self-tuning models
\cite{Fors5}.
%\newpage
%%%%%%%%%%%%%%%%%%%%%%%%%%
%%%%%%%%%%%%%%%%%%%%%%%%%%
%%%SECTION 3
\section{Choosing a Gauge}\label{sec:Gau}
As in the RS models, the inclusion of matter entails a shift of
branes from their vacuum positions - this is the basis for the
moduli space approximation. Instead, by a gauge transformation,
the branes can be restored to their vacuum coordinate locations
\cite{Char17,Brax15} at the price of introducing an $x^{\mu}$
dependence in the metric itself. It is advantageous to maintain
$g_{(5) \mu 5} = 0$ and separate the graviton zero mode $g_{\mu
\nu} (x)$.
%%%%%%%%%%%%%%%%%%%%%%%%%%%%%%%%%%
%%%NEW
At energies small compared to
the scale of $W$ we
may neglect the Kaluza-Klein modes
implicit in $g_{(5) \mu \nu}$ [18].
%%%%%%%%%%%%%%%%%%%%%%%%%%%%%%%%%%%%
%%%%%%%%%%%%%%%%%%%%%%%%%%%%%%%%%%%%
Thus we
consider the
%%%%%%%%%
%%%NEW
sufficiently general
%%%%%%%%%%%%
%%%%%%%%%%%
ansatz
%%%EQUATION 26
\vspace{-0.25cm}
\begin{equation}
d S_{(5)}^{2} = \Psi^{2} (x, y) \; g_{\mu \nu} (x) \; d x^{\mu} \;
d x^{\nu} - \varphi^{2} (x, y) \; d y^{2} \; \; .
\end{equation}
Note that
$\displaystyle{d_{5} (x) = \int_{0}^{\ell} \; \varphi (x, y) \; d y}$
measures the proper distance between
the branes at fixed $x^{\mu}$.

The Christoffel symbols, Ricci tensor and Ricci scalar for the
metric Eq. (26) have been given in \cite{Kim16}. There the metric
functions $\Psi$ and $\varphi$ are restricted by $R_{(5) \mu 5} =
0$ in the $AdS_{5}$ bulk. Here, due to the bulk scalar we rather
have
%%%EQUATION 27
\begin{equation}
\displaystyle{
G_{(5) \mu 5} = R_{(5) \mu 5}
              = 3 \left[ \left(
                \frac{\Psi '}{ \Psi}
                \right) \;
                \left( \frac{\varphi_{, \mu} }{\varphi}
                \right)
                - \left(
                \frac{ \Psi '}{ \Psi}
                \right)_{, \mu} \right]
              = K_{(5)} \; T_{(5) \mu 5}^{(\Phi)} =
\Phi_{, \mu} \; \Phi ' \; \; .
}
\end{equation}
To deal with the nonvanishing right-hand side let us
%%%%%%%%%%%%%%
%%%NEW
take\footnote{In the case of the exponential superpotential one
can find other solutions of Eq. (27) analogous to \cite{Band19}.}
%%%%%%%%%%%%%%%
%%%%%%%%%%%%%%
%%%EQ 28
\begin{equation}
\displaystyle{
\Psi (x, y) = \mbox{exp} \; \left( - A (F (x, y) ) \right) \; , \;
\Phi (x, y) = \Phi \left( F (x, y) \right)
}
\end{equation}
where $A(z)$ and $\Phi (z)$ are solutions of Eqs. (16,17). That is
to say
%%%EQ 29
\begin{eqnarray}
\begin{array}{lcl}
\Phi ' &=& \displaystyle{ \frac{1}{2} \; \frac{\partial W}{
\partial \Phi} \; F ' \; , \; \Phi_{, \mu} =
\frac{1}{2} \; \frac{\partial W}{ \partial \Phi} \;
F_{, \mu} \; , \; -
\left( \frac{\Psi '}{ \Psi} \right) = \frac{1}{6} \;
W \; F ' \; \; , } \\[.5cm]
       &-& \displaystyle{ \left( \frac{\Psi '}{ \Psi}
       \right)_{, \mu} = \frac{1}{6} \; W \; F'_{, \mu} + \frac{1}{12} \;
\left( \frac{\partial W}{ \partial \Phi } \right)^{2} \; F_{, \mu} \; F ' }
\end{array}
\end{eqnarray}
yielding
%%%EQ 30
\vspace{-0.25cm}
\begin{equation}
W \; \left[ F\;'_{, \mu} - F\;' \; \frac{\varphi_{, \mu}}{\varphi}
\right] = W \varphi \left( \frac{F'}{\varphi} \right)_{, \mu}
= 0 \; \; .
\end{equation}
Thus $F'/\varphi$ can be at most a function of $y$ only which is
fixed to unity by the vacuum, i.e.
%%%EQ 31
\begin{equation}
\varphi (x, y) = F\;' (x,y) \; \; .
\end{equation}
Eqs. (28,30) are consistent with the perturbative results
\cite{Brax15}.
%\\[.2cm]
In the terminology of \cite{Kim16} different choices of the one
free scalar function $F (x,y)$ are `gauges'.
%%%%%%%%%%%%%%%
%%%NEW
This is not to say that they necessarily describe the same
physics, however, as we will show.
%%%%%%%%%%%%
%%%%%%%%%%%%
%%%%%%%%%%%%%%%%%%%%
%%%OUT
% An obvious choice is $F
%(x,y) = \varphi (x) y$ which corresponds to the first gauge of
%Chiba \cite{Chib10} and that of Kanno and Soda \cite{Kann18} for
%the RS model. However, as discussed in \cite{Kim16}, that gauge
%has the unwanted feature that the scalar mode disappears from the
%second RS model, and the same holds true for the improved gauge at
%\cite{Char17} (second gauge of \cite{Chib10}). Essentially the
%problem arises because the starting action is in the Einstein
%frame, where perturbation theory is formulated, whereas the
%effective action is in the Jordan frame.
%\newpage
%%%%%%%%%%%%%%%%%%%%%%%%%%%%%
%\newpage
\section{The Effective Action in the Einstein Gauge}\label{sec:Ein}
For the metric Eq. (26) the bulk action of Eq. (1) is
%%%EQ 32
\begin{eqnarray}
\begin{array}{ll}
S_{\rm bulk} \; = \; &
\displaystyle{\frac{1}{K_{(5)}} \; \int \; d^{4} x \;
\sqrt{- g} \; \int \; dy \; \biggl\{
- \frac{R}{2} \; \Psi^{2} \; \varphi - 3 g^{\mu \nu} \;
\left( \varphi \Psi \right)_{, \mu} \;
\Psi_{, \nu} \biggr.}\\[.5cm]
                     &
\displaystyle{+ \; \biggl. 6 \; \frac{ \left( \Psi \Psi\;'
\right)^{2}}{\varphi} - 4 \;  \left( \frac{\Psi^{3}
\Psi\;'}{\varphi} \right)^{'} + \frac{1}{2} \;
\Psi^{2} \; \varphi \; g^{\mu \nu} \; \Phi_{, \mu}
\; \Phi_{, \nu} \biggr.}\label{Sbulk}\\[.5cm]
                     &
\displaystyle{- \biggl. \frac{1}{2} \;\frac{\Psi^{4}}{\varphi} \;
\left( \Phi\;' \right)^{2} - \Psi^{4} \; \varphi \; U (\Phi)
\biggr\} \; \; .}
\end{array}
\end{eqnarray}
Here we can omit the $y$ integral of the total derivative term
which is cancelled by the implicit Gibbons-Hawking terms. As in
the RS case \cite{Kim16}, we impose as a gauge condition that the
coefficient of the four-dimensional Ricci scalar $R$ be identical
to the vacuum solution:
%%%EQ 33
\begin{equation}
\displaystyle{ \Psi^{2} \; \varphi = {\rm e}^{- 2 A (F)} \; F\;' =
{\rm e}^{-2 A (y)} \; \; , }
\end{equation}
where we have used Eqs. (28,31). Implicitly this determines $F (x,
y)$ as
%%%EQ 34
\begin{equation}
\displaystyle{
\int_{y}^{F (x,y)} \; dz \; {\rm e}^{- 2 A (z)} + T (x) = 0 \; \; ,
}
\end{equation}
the integration function $T (x)$ being related to the radion.
Observe the physical distance between the branes is
%%%EQ 35
\vspace{-0.5cm}
\begin{equation}
\displaystyle{ d_{5} (x) = F (x, \ell) - F (x, 0) \; \; .}
\end{equation}
Per definition
%%%EQ 36
\begin{equation}
\displaystyle{
\frac{1}{K} = \frac{2}{K_{(5)}} \; \int_{0}^{\ell} \;
{\rm e}^{- 2 A (y)} \; d y
}
\end{equation}
so we may write
%%%EQ 37
\vspace{-0.25cm}
\begin{equation}
\displaystyle{
S_{\rm bulk} = \int \; d^{4} x \; \sqrt{- g} \; \Biggl\{
- \frac{R}{2K} + \cal{L}_{\rm bulk} \Biggr\}
}
\end{equation}
with, using Eqs. (15,16,17,28,31,33,34),
%%%EQUATION 38
\begin{eqnarray}
\begin{array}{lcl}
\cal{L}_{\rm bulk} & = & \displaystyle{ \frac{2}{K_{(5)}}
\; \int_{0}^{\ell} \; d y \; \Biggl\{ 3 \; {\rm e}^{- 2 A (F)} \;
F\;' \; g^{\mu \nu} \; \left( A (F) \right)_{, \mu} \;
\left( A (F) \right)_{, \nu} \Biggr. }\\[.75cm]
&   & + \; \displaystyle{ 6 \; \frac{{\rm e}^{- 4 A (F)} \left(
(F)' \right)^{2}}{F\;'} + \frac{1}{2} \; {\rm e}^{- 2 A (F)} \;
F\;' \; g^{\mu \nu} \;
\left( \Phi (F) \right)_{, \mu} \; \left( \Phi (F)
\right)_{, \nu} }\\[.75cm]
&   & \displaystyle{ - \Biggl. \; \frac{1}{2} \; \frac{{\rm e}^{-
4 A (F)}}{F\;'} \; \left( \Phi (F)' \right)^{2} - {\rm e}^{- 4 A
(F)} \; F\;' \; \biggl[ \frac{1}{8} \; \left( \frac{\partial
W}{\partial \Phi} \right)^{2} -
\frac{W^{2}}{6} \biggr] \Biggr\} } \\[.75cm]
   & = & \displaystyle{ \frac{2}{K_{(5)}}
   \; \int_{0}^{\ell} \; dy \; F\;' \; \Biggl\{
   \; 3\;{\rm e}^{- 2 A (F)} \; g^{\mu \nu} \;
   \left( - \frac{W}{6} \; {\rm e}^{2 A(F)} \; T_{, \mu}
   \right) \; \left( - \frac{W}{6} \;
          {\rm e}^{2 A(F)} \; T_{, \nu} \right) \Biggr.  }\\[.75cm]
     & & + \; \displaystyle{ 6\;{\rm e}^{- 4A (F)} \;
     \left( \frac{W}{6} \right)^{2} + \frac{1}{2} \; {\rm e}^{- 2 A
(F)} \; g^{\mu \nu} \; \left( - \frac{1}{2} \; \frac{\partial
W}{\partial \Phi} \;
          {\rm e}^{2 A (F)} \; T_{, \mu} \right) \cdot }\\[.75cm]
     & & \displaystyle{ \left( - \frac{1}{2} \;
     \frac{\partial W}{\partial \Phi} \;
     {\rm e}^{2 A (F)} \; T_{, \nu} \right) \;
          - \; \frac{1}{2} \; {\rm e}^{- 4 A (F)} \;
          \left( \frac{1}{2} \; \frac{\partial W}{\partial \Phi}
           \right)^{2}   }\\[.75cm]
      & & \displaystyle{ \Biggl. - \; {\rm e}^{- 4 A (F)}
      \; \biggl[ \frac{1}{8} \; \left( \frac{\partial W}{\partial \Phi}
      \right)^{2} -   \frac{W^{2}}{6} \biggr] \Biggr\} }\\[.75cm]
%\end{array}
%\end{eqnarray*}
%\begin{eqnarray}
%\begin{array}{ll}
            & = & \displaystyle{\frac{2}{K_{(5)}}
\; \int_{F (x, 0)}^{F (x, \ell)} \; dz \;
\Biggl\{ \frac{1}{2} \; {\rm e}^{2A (z)} \; \biggl[
\frac{W^{2}}{6} + \frac{1}{4} \;
\left( \frac{\partial W}{\partial \Phi} \right)^{2}
\biggr] \; g^{\mu \nu} \; T_{, \mu} \; T_{, \nu} \Biggr.  }\\[.75cm]
   &  & \displaystyle{ \Biggl. + \; {\rm e}^{- 4 A (z)} \; \left[
\frac{W^{2}}{3} - \frac{1}{4} \; \left( \frac{\partial W}{\partial
\Phi} \right)^{2} \right] \Biggr\} \; \; .}
\end{array}
\end{eqnarray}
\noindent
Here $W = W ( \Phi (z) )$ and similarly for $\partial W/
\partial \Phi$.
%%%%%%%%%%%%%%%%
%%%NEW
As $2 \left( \partial W / \partial \Phi \right)^{2}
= d W / d z$ the integral becomes a total derivative,
yielding
%%%EQ(39)
\begin{eqnarray}
%\begin{array}{lcl}
{\cal{L}}_{\rm bulk}  =  \displaystyle{
g^{\mu \nu} \; T_{, \mu} T_{, \nu} \;
\left[ \frac{W \rm{e}^{2A}}{2 K_{(5)}} \right]_{F(x, 0)}^{F(x, \ell)}
- \left[ \frac{W \rm{e}^{-4A}}{K_{(5)}} \right]_{F(x, 0)}^{F(x, \ell)} } \; \; .
\end{eqnarray}
The latter terms in Eq.(39) cancel with the brane
potentials as expected since the vacuum solution does
not admit a net cosmological constant.
%%%%%%%%%%%%%%%%%
%%%%%%%%%%%%%%%%%
%%%%%%%%%%%%%%%%%%%%%%%%%%%%
%%%OUT
%\frac{2}{K_{(5)}}
%\; \int_{0}^{\ell} \; d y \; \Biggl\{ 3 \; {\rm e}^{- 2 A (F)} \;
%F\;' \; g^{\mu \nu} \; \left( A (F) \right)_{, \mu} \;
%\left( A (F) \right)_{, \nu} \Biggr. }\\[.75cm]
%Eqs. (34,36,39) constitute our general result for BPS branes.
%\\[.2cm]
As an example, we obtain for the experimental superpotential of
Eq. (19), using Eqs. (24, 36)
%%% EQUATION 40
\begin{equation}
\frac{1}{K} = \frac{12 ( 1 - \beta)}{ \left( 2 + 3 \alpha^{2} \right)
K_{(5)} W_{0} } \; , \; \;
\beta \equiv \left( 1 - \frac{\alpha^{2}}{2} \; W_{0} \ell
\right)^{ 1 + \frac{2}{3 \alpha^{2}} } \; \; .
\end{equation}
and by Eq. (34)
%%% EQ 41
\begin{eqnarray}
\begin{array}{lcl}
\displaystyle{ \biggl[ 1 - \frac{\alpha^{2}}{2} \; W_{0} \; F (x,
y ) \biggr]^{1 + \frac{2}{3 \alpha^{2} } } } &  -    &
\displaystyle{ \left( 1 - \frac{\alpha^{2}}{2} \; W_{0} \; y
\right)^{1 + \frac{2}{3 \alpha^{2} } } }
   =
\displaystyle{
\frac{ \left( 2 + 3 \alpha^{2} \right) }{6} \; W_{0}
\; T (x)  } \\[.5cm]
& \equiv & \displaystyle{ \phi (x) } \; \; .
\end{array}
\end{eqnarray}
%%%%%%%%%%%%%%%%%
%%%%%%%%%%%%%%%%
%%%NEW
The complete effective action in the Einstein gauge is
\begin{equation}
S_{\rm eff} = \int \; d^{4} x \; \sqrt{-g} \;
\left[ - \frac{R}{2K} + \frac{\omega ( \phi )}{2K} \; g^{\mu \nu} \;
\phi_{, \mu} \; \phi_{, \nu} \right] + S^{(+)} + S^{(-)}
\end{equation}
where
%The remaining integrals in Eq. (39) are straightforward with the
%result
%%%EQ 42
%\begin{equation}
%{\cal{L}}_{\rm bulk} = \displaystyle{ \frac{\omega ( \phi)}{2 K}
%\; g^{\mu \nu} \; \phi_{, \mu} \; \phi_{, \nu} +
%\frac{1}{K_{(5)}} \;
%\biggl[ \Psi^{4} (x, 0) \; W \left( \Phi (x, 0) \right) -
%       \Psi^{4} (x, \ell) \; W \left( \Phi (x, \ell) \right)
%       \biggr] \; \; , }
%\end{equation}
%%%EQ 43
\begin{equation}
\omega ( \phi) \; = \;
\displaystyle{
\frac{3}{2 + 3 \alpha^{2} }
\; \frac{1}{ (1 + \phi)(\beta + \phi)}
} \;  \;.
\end{equation}
$S^{(+)}$ describes matter on the positive tension brane at
$y = 0$ which feels the metric $g_{ \mu \nu}^{(+)} (x) \;
= \; g_{(5) \mu \nu}$
$(x, 0) = \Psi^{2} (x, 0) \; g_{\mu \nu} (x)$,
%
% \; \; \beta = \left(
%1 - \frac{\alpha^{2}}{2} \; W_{0} \ell \right)^{1 + \frac{2}{3
%\alpha^{2}}} \; \; .}
%\end{equation}
%The last two terms in Eq. (42) cancel the brane potentials, as
%expected, since no net cosmological constant is admitted.
%\\[.2cm]
%Matter on the brane at $y = 0$ feels the metric
%$\displaystyle{g_{(5) \mu \nu} \; (x, 0) = \Psi^{2} (x, 0)
%\; g_{\mu \nu} (x)}$,
%%%EQ 44
\begin{equation}
\displaystyle{ \Psi^{2} (x, 0) = \biggl[ 1 + \phi (x)
\biggr]^{\frac{2}{2 + 3 \alpha^{2}}}  }
\end{equation}
and in the Horava-Witten model also couples to
%%%EQ 45
\begin{equation}
\displaystyle{ W \left( \Phi (x, 0) \right) / W_{0} = \biggl[ 1 +
\phi (x) \biggr]^{-\frac{3 \alpha^{2}}{2 + 3 \alpha^{2}}} \; \; .}
\end{equation}
On the negative tension brane at $y = \ell$ corresponding to $S^{(-)}$
the conformal factor is
%%% EQ 46
\begin{equation}
\displaystyle{ \Psi^{2} (x, \ell) = \biggl[ \beta + \phi (x)
\biggr]^{\frac{2}{2+3 \alpha^{2}}}  }
\end{equation}
and
%%%EQ 47
\begin{equation}
\displaystyle{ W \left( \Phi (x, \ell) \right) / W_{0} = \biggl [
\beta + \phi (x) \biggr]^{-\frac{3 \alpha^{2}}{2+3 \alpha^{2}}}\;
\; .}
\end{equation}
In the limit $\alpha \rightarrow 0$, Eqs. (42,43,44,46) go
smoothly to the expression for the RS models in \cite{Kim16}.
Irrespective of the value of $\alpha$, or whether $\Phi$ couples
directly to matter, $\beta$ is the key parameter controlling the
strength of the scalar coupling: if the coordinate position of the
second brane is sufficiently close to the bulk singularity one can
satisfy the constraints on scalar-tensor theories \cite{Will20},
and the Horava-Witten value $\alpha = 1$ is cosmologically safe
up to the issue of nonrelativistic matter \cite{Arno12}.

%%%%%%%%%%%%%%%%%%%%%%%%%%%%%%
%%%%%%%%%%%%%%%%%%%%%%%%%%%%%%
%%%START SECTION 5 HERE
%%%%%%%%%%%%%%%%%%%%%%%%%%%%%%
%%%%%%%%%%%%%%%%%%%%%%%%%%%%%%
\section{The Jordan and Moduli Space Gauges}\label{sec:Jor}
Often in scalar-tensor theories one gives priority to the Jordan
frame, in which the motion of feducial test particles is geodesic,
rather than the Einstein frame where the scalar and tensor fields
are unmixed. One can impose the Jordan frame on the positive
tension brane as a gauge condition by taking
%%%EQ 48
\begin{equation}
F (x, y) = y \varphi (x)
\end{equation}
as in \cite{Chib10,Kann18} for the RS model.\footnote{See,
however, \cite{Char17} for some cautionary remarks.} Proceeding
from Eq. (32) in this Jordan gauge a straightforward calculation
now yields
%For the exponential super potential, using Eq.(24)
%%%EQ 49
\begin{eqnarray*}
%\begin{array}{lcl}
S_{\rm bulk}  =  \displaystyle{
\int \; d^{4} x \; \sqrt{-g} \; \left\{  -  \frac{R}{K_{(5)}} \;
\int_{0}^{\ell \varphi} \; dz \; {\rm e}^{- 2 A (z)} \; + \;
\frac{g^{\mu \nu}}{2 K_{(5)}} \right.
} \;
\displaystyle{
\left[ W \; {\rm{e}}^{-2A} \; z_{, \mu} \; z_{, \nu}
\right]_{0}^{\ell \varphi}
}
\end{eqnarray*}
\begin{eqnarray}
%\displaystyle{
    -  \; \left.
\left[ \frac{W}{K_{(5)}} \; {\rm{e}}^{-4A}
\right]_{0}^{\ell \varphi} \right\}
%} \; \; .
\end{eqnarray}
For the exponential super potential, using Eq. (24)
%%%EQ 50
\begin{eqnarray}
\begin{array}{lcl}
\displaystyle{ \frac{1}{K_{(5)}} \; \int_{0}^{\ell \varphi}dz \;
{\rm e}^{- 2 A (z)} } & =      & \displaystyle{ \frac{6}{K_{(5)}
\; W_{0} (2 + 3 \alpha^{2})} } \; \displaystyle{ \left[ 1 - \left(
1 - \frac{\alpha^{2}}{2} \; W_{0} \; \ell \varphi \right)^{1 + 2/3
\alpha^{2} } \right]
}  \\[.5cm]
& \equiv &
\displaystyle{
\frac{1}{2 \; K_{0} } \; \left[ \psi \right]
}
\end{array}
\end{eqnarray}
where $\psi$ is the Brans-Dickie scalar and $K_{0}$ a bare gravitational coupling.
Including matter, the effective action is then
%%%EQ 51
\begin{equation}
S_{\rm eff} = \int \; d^{4} x \; \sqrt{-g} \;
\left[ - \frac{\psi R}{2K_{0}} + \frac{\omega ( \psi )}{2K_{0}} \; g^{\mu \nu} \;
\psi_{, \mu} \; \psi_{, \nu} \right] + S^{(+)} + S^{(-)}
\end{equation}
%%%EQ 52
\begin{equation}
\omega (\psi) \; = \; \frac{3}{(2 + 3 \alpha^{2}) (1 - \psi)} \; \; .
\end{equation}
Note $\omega (\psi)$ drives $\psi$ to unity by the self-tuning mechanism \cite{Chib10}.
In this gauge matter on the negative tension experiences a metric
%%%EQ 53
\begin{equation}
g_{\mu \nu}^{(-)} (x) \; = \; \left( 1 - \psi \right)^{\frac{2}{2 +
3 \alpha^{2}} } \; g_{\mu \nu} (x)
\end{equation}
and can be coupled to
%%%EQ 54
\begin{equation}
W \; \left( \Phi ( \ell \varphi) \right) / W_{0} \; = \;
\left( 1 - \psi \right)^{- \frac{3 \alpha^{2} } {2 + 3 \alpha^{2} } } \; \; .
\end{equation}
Per the gauge definition, $g_{\mu \nu}^{(+)} (x) = g_{\mu \nu}
(x)$ on the positive tension brane, and moreover $W = W_{0}$ there
which is to say the matter is implicitly decoupled from the bulk
scalar. Nor, for that matter, can radiation-scalar coupling be
recovered by a conformal transformation of the Jordan gauge
effective action to the Einstein frame $g_{\mu \nu} \rightarrow
g_{\mu \nu} / \psi$. This makes the Jordan gauge unsuitable for
the Horava-Witten model.

 Within a given scalar-tensor theory the
Jordan and Einstein frames describe identical physics, but the
Jordan and Einstein gauges are inequivalent even in the absence of
direct coupling of the bulk scalar and brane matter. The
coordinate length $\ell$ appears directly in the Einstein gauge
via the coupling parameter $\beta$, whereas in the Jordan gauge it
is subsumed in $\psi$. The two gauges only become conformally
equivalent if $\alpha = \beta = 0$. In that case $\psi = (1 +
\phi)^{-1} = 1 - \frac{\chi^{2}}{6}$ with $\chi$ a conformally
coupled scalar \cite{Fadd21}.

 Still,
there is a subtlety with $\alpha = \beta = 0$ : $\alpha
\rightarrow 0$ followed by $\ell \rightarrow \infty, \beta
\rightarrow 0$ in the Einstein gauge describes the second RS
model. Although the coordinate distance is infinite, the AdS warp
makes the physical distance finite. Displacing the brane distorts
the bulk geometry as reflected in the scalar $\phi$ remaining in
the effective action \cite{Kim16}. Taking $\alpha \rightarrow 0$
followed by $\ell \rightarrow \infty$ in the Jordan gauge would
yield $\psi = 1$, according to Eq. (50), and no scalar which is
the wrong physics.

 Next, we turn to the moduli space approximation. In
the original version \cite{Brax13} (see also \cite{Khou22}) the
Minkowski metric of the static vacuum solution in Section
\ref{sec:BPS} is promoted to $g_{\mu \nu} (x)$ and the brane
positions to $X^{(\pm) 5} (x)$ with $h_{\mu \nu}^{(\pm)} (x) =
g_{(5) \mu \nu} \left( x, X^{(\pm) 5} \right) - X_{, \mu}^{(\pm)
5} \; X_{, \nu}^{(\pm) 5}$, the induced metrics on the branes. The
alternative formulation of \cite{Palm14} is equivalent to here
imposing a moduli space gauge
%%%EQ 55
\begin{equation}
F(x, y) = \varphi (x) y - \xi (x) \; \; .
\end{equation}
The additional field $\xi$ represents the centre of mass
displacement, or a local twist of the orbifold boundary conditions
\cite{Lehn23}. Once again a straightforward calculation proceeding
from Eq.(32) yields\footnote{The precise form in \cite{Palm14}
obtains by a conformal transformation $g_{\mu \nu} \rightarrow
{\rm exp}(2 A (z^{+}) ) \; g_{\mu \nu}$ to the Jordan frame on the
positive tension brane.}
%%%%EQ 56
\begin{eqnarray*}
%\begin{array}{lcl}
S_{\rm bulk}  =  \displaystyle{
\int \; d^{4} x \; \sqrt{-g} \; \left\{  -  \frac{R}{K_{(5)}} \;
\int_{z^{+}}^{z^{-}} \; dz \;\; {\rm e}^{- 2 A} \; + \;
\frac{g^{\mu \nu}}{2 K_{(5)}} \right.
} \;
\displaystyle{
\biggl[ W  \; {\rm{e}}^{-2A}   \;
z_{, \mu} \; z_{, \nu}
\biggr.
}
\displaystyle{
\biggl. \biggr]_{z^{+}}^{z^{-}}
}
\end{eqnarray*}
\begin{eqnarray}
%\left.  -  \;
\displaystyle{
-  \left. \left[ \frac{W}{K_{(5)}} \; \; \;
{\rm{e}}^{-4A} \right]_{z^{+}}^{z^{-}} \right\}
} \; \; .
\end{eqnarray}
Evidently the moduli space gauge $S_{\rm bulk}$, Eq. (56), is just
two copies of the Jordan gauge $S_{\rm bulk}$, Eq. (49), glued
together. The Einstein gauge does not allow the shift mode $\xi$
but one could paste together two copies with scalars $\phi^{(+)}$
and $\phi^{(-)}$. The key point is that where the joint is made
one must impose Israel's junction conditions - e.g. at $z = 0$
\begin{equation}
\left[ \left[ A' \right] \right] = 0 \;\;\; , \;\;\;
\left[ \left[ \Phi' \right] \right] = 0
\end{equation}
where $\left[ \left[ \;\; \right] \right]$ denotes the
discontinuity. One recognises Eq. (57) as the junction conditions
for a tensionless brane. This is not mere tautology: to discuss
the centre-of-mass, as opposed to relative, motion of the positive
and negative tension branes requires a third observer brane. The
catch is that Eq. (57) is not BPS unless $U ( \Phi )$ has a
zero\footnote{That is the case examined in \cite{Lehn23}.} or the
superpotential is a constant. If $W$ is constant, one has
AdS$_{5}$,
 the RS model, and two conformally coupled scalars $\chi^{(+)}, \chi^{(-)}$,
 and through a conformal transformation only one scalar mode $\left( \chi^{(-)}
 / \chi^{(+)} \right)$. Otherwise one is not examining the advertised
 two-brane BPS system, but instead a nearly BPS three-brane system
 similar to the ekpyrotic model of \cite{Lehn23}.
%%%%%%%%%%%%%%%%%%%%%%%%%%%%%%%%%%%%%%%%%%
%%%%%OUT
%One can now readily understand what underlies of the conclusions of \cite{Brax13}.
%The centre-of-mass mode is not %self-tuning because there is no restoring force.
 %In the analogue picture of the orbifold offered in \cite{Lehn23} the
 %child's climbing toy unwinds. Like the apparent disappearance of
 %the radion in the second RS model this is the wrong %physics.
%%%%%%%%%%%%%%%%%%%%%%%%%%%%%%%%%%%%%%%%%%
%%%NEW REPLACING LAST PARAGRAPH 7/3/05

The ramifications of the moduli space gauge becomes evident
by adapting the viewpoint of a freely falling observer on the
tensionless brane rather than the positive tension brane. For
the exponential superpotential define
%%%EQ 58
\begin{equation}
\left( 1 - \frac{\alpha^{2}}{2} \;\; W_{0} \; z^{+}
 \right)^{1 + \frac{2}{3 \alpha^{2}}}
\; = \;
\psi \; \cosh^{2} \;
\left( \frac{r}{2} \right)
\end{equation}
%%%EQ 59
\begin{equation}
\left( 1 - \frac{\alpha^{2}}{2} \;\; W_{0} \; z^{-} \right)^{1 +
 \frac{2}{3 \alpha^{2}}}
\; = \;
\psi \; \sinh^{2} \;
\left( \frac{r}{2} \right)
\end{equation}
and ignore brane matter so
%%%EQ 60
\begin{equation}
S_{\rm eff} = \int \; d^{4} x \; \sqrt{-g} \;
\left\{ - \frac{R \psi}{2K_{0}} + \frac{g^{\mu \nu}}{2K_{0}} \;
\left( \frac{3}{2 + 3 \alpha^{2}} \right) \;
\left[ \frac{\psi_{, \mu} \; \psi_{, \nu} }{ \psi} - \psi \;
r_{, \mu} \; r_{, \nu} \right] \right\} \; \; .
\end{equation}
Such observers see a Brans-Dickie theory coupled to a ghost. The instability
reflects that the tensionless brane wants to sit at $z = - \infty$ where $U$
vanishes. Like the apparent disappearance of the radion in the RS2 limit,
this is the wrong physics.
%\section{Conclusions}
%BPS braneworlds are much closer to string/$M$-theory than the
%simple RS models. Our general Einstein gauge effective action, Eqs. (34, 37, 39)
%provide a simple treatment of the nonvacuum case without invoking
%the third brane of the moduli space approximation. It can
\section{Conclusions}\label{sec:Con}
BPS braneworlds are much closer to string/$M$-theory than the
simple RS models. Our general Einstein gauge effective action,
Eqs. (34, 37, 39) provide a simple treatment of the nonvacuum case
without invoking the problematic third brane of the moduli space
approximation. It can be used even when there is direct coupling
of the bulk scalar and brane matter, unlike the Jordan gauge. The
metric of Eq. (26) could be used as a starting point for
calculating Kaluza-Klein corrections through the low energy
expansion scheme \cite{Kann18}. Setting $\alpha = 1$ in Eqs.
(42-47) one has a new basis to explore inhomogeneous Horava-Witten
brane cosmology and, following the methods of \cite{Fadd21}, also
black holes.

%\begin{appendix}
\section*{Acknowledgements}
One of us (J.E.K.)  is supported in part by the KOSEF Sundo Grant,
the ABRL Grant No. R14-2003-012-01001-0, and the BK21 program of
Ministry of Education, Korea. Two of us (R.D.V. and G.B.T.)
acknowledge grants from the South African National Research
Foundation (NRF GUN-2053794), the Research Committee of the
University of Cape Town and the Foundation for Fundamental
Research (FFR PHY-99-01241).
\newpage
\section*{Appendix A - Non BPS Branes}
Following \cite{Kann9}, suppose the bulk potential is
\begin{center}
%\begin{eqnarray*}
$\displaystyle{U ( \Phi ) \; = \; u( \Phi) - 6 \; k^{2}}$ \marginpar{(A.1)}
%\end{eqnarray*}
\end{center}
and on the positive tension brane at $y = 0$
%\begin{equation}
\begin{center}
$\displaystyle{V_{0} ( \Phi ) \; = \; 6 k + v_{0} ( \Phi )}$ \marginpar{(A.2)}
\end{center}
%\end{equation}
while on the negative tension brane at $y = \ell$
\begin{center}
%\begin{equation}
$\displaystyle{V_{\ell} ( \Phi ) \; = \; - 6 k + v_{\ell} ( \Phi )}$ \marginpar{(A.3)}
%\end{equation}
\end{center}
with $u$, $v_{0}$ and $v_{\ell}$ small. Neglecting the influence
of the scalar on the bulk geometry one can replace $\Phi (x, y)$
with the zero mode $\eta (x)$ at leading order. The metric
functions for the RS geometry are \cite{Kim16}
%\begin{equation}
\begin{center}
$\displaystyle{\Psi (x, y) \; = \; \bigg[ {\rm e}^{- 2 k y}
+ \phi (x) \biggr]^{1/2} \; , \;
\varphi (x, y) = \Psi^{-2} (x, y) \; {\rm e}^{-2 k y} \; \;
. }$ \marginpar{(A.4)}
%\end{equation}
\end{center}
A simple calculation using Eqs. (32,36) gives
\begin{center}
%\begin{eqnarray}
$\displaystyle{
\begin{array}{ll}
{\cal{L}} \; = \; & \displaystyle{ - \frac{R}{2K} + \frac{3}{4K} \;
\frac{ g^{\mu \nu} \; \phi_{, \mu} \; \phi_{, \nu} }
     { (1 + \phi) \left( e^{- 2 k \ell} + \phi \right) }
     + \frac{ g^{\mu \nu} }{2K} \; \eta_{, \mu}
      \; \eta_{, \nu} }\\[.75cm]
                  & - \; \displaystyle{ \frac{1}{K}
      \left( \frac{ 1 + e^{- 2 k \ell}}{2} + \phi \right) \;
       u ( \eta) - \frac{k}{K \left( 1 - e^{- 2 k \ell} \right) } \;
       \biggl[ \left( 1 + \phi \right)^{2} \; v_{0} ( \eta )
       \biggr. }\\[.75cm]
       & + \; \displaystyle{ \biggl. \left( {\rm e}^{- 2
       k \ell} + \phi \right)^{2} \; v_{\ell} (\eta) \biggr] } \; \; .
\end{array} }$ \marginpar{(A.5)}
%\end{eqnarray}
\end{center}
Note the radion $\phi$ remains in the limit $\ell \rightarrow \infty$.

%\end{appendix}
%%%%%%%%%%%%%%%%%%%%%%%%%%%%%%%%%%%%%%%%%%%%%%%%%%%
%%%%%%%%%%%%%%%%%%%%%%%%%%%%%%%%%%%%%%%%%%%%%%%%%%
%%%THE END
%%%%%%%%%%%%%%%%%%%%%%%%%%%%%%%%%%%%%%%%%%%%%%%%%%%%


\newpage

\begin{thebibliography}{99}

\def\apj#1#2#3{Astrophys.\ J.\ {\bf #1}, #2 (#3)}
\def\ijmp#1#2#3{Int.\ J.\ Mod.\ Phys.\ {\bf #1}, #2 (#3)}
\def\mpl#1#2#3{Mod.\ Phys.\ Lett.\ {\bf A#1}, #2 (#3)}
\def\npb#1#2#3{Nucl.\ Phys.\ {\bf B#1}, #2 (#3)}
\def\plb#1#2#3{Phys.\ Lett.\ {\bf B#1}, #2 (#3)}
\def\prd#1#2#3{Phys.\ Rev.\ {\bf D#1}, #2 (#3)}
\def\prl#1#2#3{Phys.\ Rev.\ Lett.\ {\bf #1}, #2 (#3)}
\def\prt#1#2#3{Phys.\ Rep.\ {\bf #1}, #2 (#3)}
\def\rmp#1#2#3{Rev. Mod. Phys.\ {\bf #1}, #2 (#3)}
\def\sjnp#1#2#3{Sov.\ J.\ Nucl.\ Phys.\ {\bf #1}, #2 (#3)}
\def\zp#1#2#3{Z.\ Phys.\ {\bf #1}, #2 (#3)}
\def\jhep#1#2#3{JHEP\ {\bf #1}, #2 (#3)}
\def\ephjc#1#2#3{Europhys. J. C\ {\bf #1}, #2 (#3)}

\bibitem{Maa1}
R. Maartens, {\it Living Rev. Rel.} {\bf 7}, 1 (2004);
Ph. Brax, C. van de Bruck and
A.C. Davis, hep-th/0404011.
\bibitem{RS2}
L. Randall and R. Sundrum, \prl{83}{3370}{1999}
\prl{83}{4690}{1999}.
\bibitem{Gold3}
W.D. Goldberger and M.B. Wise,
\prl{83}{4922}{1999}.
\bibitem{Ark4}
N. Arkani-Hamed, S. Dimopoulos, N. Kaloper and R. Sundrum,
\plb{480}{193}{2000};
S. Kachru, M. Schulz and E. Silverstein,
\prd{62}{045021}{2000}.
\bibitem{Fors5}
S. Forste, Z. Lalak, S. Livagnac and H.P. Nilles,
\plb{481}{360}{2000}.
\bibitem{Koba6}
S. Kobayashi, K. Koyama and J. Soda,
\plb{501}{157}{2001};
Y. Himemoto and M. Sasaki,
\prd{63}{044015}{2001}.
\bibitem{Luka7}
A. Lukas, B. Ovrut, K. Stelle and D. Waldran,
\npb{552}{246}{1999}, Phys. Rev. {\bf D59}, 086001 (1999).
\bibitem{Hora8}
P. Horava and E. Witten,
\npb{460}{506}{1996},
\npb{475}{94}{1996}.
\bibitem{Kann9}
S. Kanno and J. Soda,
Gen. Rel. Grav. {\bf 36}, 689 (2004).
\bibitem{Chib10}
T. Chiba,
\prd{62}{021502}{2000}.
\bibitem{Dewo11}
O. De Wolfe, D.Z. Freedman, S.S. Gubser and A. Karch,
\prd{62}{046008}{2000}.
\bibitem{Arno12}
R. Arnowitt, J. Dent and B. Dutta, hep-th/0405050.
\bibitem{Brax13}
Ph. Brax, C. van der Bruck, A.C. Davis and C.S. Rhodes,
\prd{67}{023512}{2003}; S.L. Webster and A.C. Davis, hep-th/0410042.
\bibitem{Palm14}
G.A. Palma and A.C. Davis, Phys. Rev. {\bf D70}, 064021 (2004), hep-th/0407036.
\bibitem{Brax15}
Ph. Brax, C. van der Bruck, A.C. Davis and C.S. Rhodes,
\plb{531}{135}{2002};
\prd{65}{121501}{2002}.
\bibitem{Kim16}
J.E. Kim, G.B. Tupper and R.D. Viollier,
\plb{593}{209}{2004}.
\bibitem{Char17}
C. Charmousis, R. Gregory and V.A. Rubakov,
\prd{62}{067505}{2000}.
\bibitem{Kann18}
S. Kanno and J. Soda,
\prd{66}{0433526}{2002},
\prd{66}{083506}{2002}.
\bibitem{Band19}
M. Bander,
\prd{69}{043505}{2004}.
%NB: hep-th/0308125 ``Expanding Cosmologies in Brane Geometries''.
\bibitem{Will20}
C.M. Will, {\it ``Theory and Experiment in Gravitational Physics''}
(Cambridge University Press, Cambridge, 1993).
\bibitem{Fadd21}
P.L. McFadden and N.G. Turok,
hep-th/0412109.
\bibitem{Khou22}
J. Khoury, B.A. Ovrut, P.J. Steinhardt and N. Turok,
\prd{64}{123522}{2001}.
\bibitem{Lehn23}
J.-L. Lehners, P. Smyth and K.S. Stelle,
hep-th/0501212.
%\bibitem{Will20}
%C.M. Will, {\it ``Theory and Experiment in Gravitational Physics''}
%(Cambridge University Press, Cambridge, 1993).
\end{thebibliography}
\end{document}